\documentclass[journal=ancac3,manuscript=article,maxauthors=10]{achemso}

\usepackage{amssymb,amsfonts,amsmath}
\usepackage[sort&compress]{natbib}

\usepackage[version=3]{mhchem} 

\author{Sergey Pogodin}
\affiliation[Universitat Rovira i Virgili]{Departament
d'Enginyeria Quimica, Universitat Rovira i Virgili 26 Av. dels
Paisos Catalans, 43007 Tarragona, Spain}
\author{Vladimir A. Baulin}
\affiliation[ICREA]{ICREA, 23 Passeig Lluis Companys, 08010
Barcelona, Spain} \alsoaffiliation[Universitat Rovira i
Virgili]{Departament d'Enginyeria Quimica, Universitat Rovira i
Virgili 26 Av. dels Paisos Catalans, 43007 Tarragona, Spain}
\email{vladimir.baulin@urv.cat} \phone{+34 977 55 85 77} \fax{+34
977 55 96 21}

\title{Can a Carbon Nanotube Pierce through a Phospholipid Bilayer?}

\begin{document}

\begin{abstract}
Great efficiency to penetrate into living cells is attributed to carbon
nanotubes due to a number of direct and indirect observations of carbon
nanotubes inside the cells. However, a direct evidence of physical
translocation of nanotubes through phospholipid bilayers and the exact
microscopic mechanism of their penetration into cells are still lacking. In
order to test one of the inferred translocation mechanisms, namely the
spontaneous piercing through the membrane induced only by thermal motion, we
calculate the energy cost associated with the insertion of a carbon nanotube
into a model phospholipid bilayer using the Single Chain Mean Field theory
which is particularly suitable for the accurate measurements of equilibrium
free energies. We find that the energy cost of the bilayer rupture is quite
high compared to the energy of thermal motion. This conclusion may
indirectly support other energy dependent translocation mechanisms such as,
for example, endocytosis.
\end{abstract}



Phospholipid bilayers are designed by Nature to protect the interior of
cells from the outside environment \cite{Alberts,Yeagle}. Despite the
weakness of hydrophobic interactions \cite{Tanford} that drive the
self-assembly of phospholipids into bilayers, cell membranes represent a
serious protective barrier \cite{Yeagle,Cooper} for external molecules,
proteins, particles as well as artificial polymers and drugs. This barrier
is quite efficient in protecting the interior of the cells and it is a
challenging task to design nano-objects that can penetrate through the
phospholipid bilayer without damaging its structure \cite%
{KleinAMP,Kelley,Murray,Tsunoda}. If such nano-objects can be tailored, they
can potentially be used for transmembrane delivery of active components into
cells \cite{Stayton}. However, the major difficulty lies in getting a direct
microscopic information regarding the interaction of nano-objects with
phospholipid layers at the molecular level in order to provide for physical
mechanisms of nano-particle translocation across the membranes. In the
absence of such microscopic information it is quite difficult to validate or
distinguish between inferred translocation mechanisms that were proposed to
explain the internalization by cells of several nano-objects such as, for
example, cell penetrating peptides \cite{Hancock1,Brogden,Hancock2,Wagner}.

Similarly, a consensus has not yet been reached regarding the translocation
mechanism of carbon nanotubes through cell membranes. Single-walled carbon
nanotubes (SWNTs) have been found inside the cells both in direct imaging
experiments using electron microscope \cite{Porter1,Porter2}, spectroscopy
\cite{Biris} and fluorescent microscopy studies \cite%
{Weisman,Dai3,Dai1,Dai2,Dai4,Bianco1,Bianco2,Fang,Ke,Raffa}. Such
experiments suggest that carbon nanotubes can efficiently penetrate into the
cells, but very little can be said about the pathway and the entry mechanism
regulating their internalization. Experimental efforts aimed to distinguish
active and passive uptake include comparison of nanotubes internalization by
living and dead cells \cite{Dai5}, change of the rate of active biological
processes by temperature control \cite{Dai1}, addition of selective chemical
agents that inhibit active uptake \cite{Dai2}, and observing nanotubes on
the cell membranes by Atomic Force Microscopy (AFM) \cite{Ebner}.

However, most of these experiments showing the translocation of carbon
nanotubes, refer to biological membranes, which have complex structure,
their properties depend on composition and many external parameters.
Biological membranes may undergo several energy consuming processes such as
endocytosis \cite{Stenmark} or phase transitions \cite{Yeagle}. Thus, it is
quite difficult to identify from these experiments a unique translocation
mechanism and discard all other possibilities by studying a particular
biological system. From this respect, the microscopic mechanism of nanotube
translocation through the membrane remains an open question.

A study of a model system of a phospholipid bilayers without inclusions and
comprising of one type of phospholipids may shed light on the plausibility
of the spontaneous translocation of carbon nanotubes \textit{via} rupture
and diffusion of the phospholipid bilayer. If we assume that the driving
force for the physical translocation of freely moving SWNTs through the
homogeneous phospholipid bilayer is indeed thermal motion, then the energy
barrier represented by the phospholipid bilayer should be comparable with
the energy of thermal motion. Apparently, the energy barrier depends on the
orientation of the SWNT with respect to the bilayer. The minimum energy
would correspond to the positions inducing less perturbation to the
phospholipid bilayer. Since in the perpendicular orientation to the bilayer
plane the SWNT interacts with a minimal number of phospholipids, this
orientation would have the minimal energy barrier. Other orientations would
induce more perturbations to the bilayer and thus would require higher
energies. Hence, an accurate estimation of the energy cost of the SWNTs
perpendicular translocation through the phospholipid bilayer would allow to
conclude about the possibility of such mechanism.

Calculation of equilibrium energies in computer simulations, that provide a
microscopic information \cite%
{Lipowsky,Tieleman,Damodaran,Damodaran1,Berkowitz,Hann,
Heller,Tuckerman,Lyubartsev,Hyvonen} related to the structure and the
dynamics of phospholipid bilayers is not straightforward, since the
simulations usually deal with a limited number of molecules in the
simulation box subject to fluctuations as well as due to the absence of a
suitable reference state \cite{Muller1,Muller3}.

Alternative to computer simulations is the use of mean field type theories
that do not include fluctuations and give direct access to the equilibrium
free energies as a result of the minimization of the free energy functional.
The Single Chain Mean Field (SCMF) theory \cite%
{Ben-Shaul1,Ben-Shaul2,Bonet1,Bonet2} is particulary suitable for such
purposes and, in addition to the speed of calculation and high capability of
parallelization, it can provide the microscopic details with the accuracy,
competing with coarse-grained Monte Carlo (MC) and Molecular Dynamics (MD)
simulations.

The numerical implementation of the SCMF theory was recently improved and
the method was applied to model the equilibrium properties of the DMPC
phospholipid bilayers in a fluid phase in Ref. \citenum{Pogodin}. We use the
3-beads model of DMPC phospholipids of this work to model fluid phase of
phospholipid bilayer and estimate the free energy of equilibrium insertion
of the SWNT in a perpendicular direction to the bilayer plane using
different diameters of carbon nanotubes and different parameters of
interaction between nanotubes and the core of the bilayer. This gives the
microscopic information about the possibility of spontaneous translocation
of SWNTs through the phospholipid bilayer by thermal motion. Our model study
has a broader interest than only study of biologically relevant
translocation mechanisms. For example, AFM tip associated with a carbon
nanotube can be used in order to quantify the force of the membrane rupture
\cite{Macpherson,Moudgil,Butt,Bau}.

\section{Free Energy of the Nanotube Insertion}

The Single Chain Mean Field (SCMF) theory is relatively fast and stable tool
for obtaining equilibrium properties and free energies of soft matter
systems with different geometries and molecular structures \cite{Pogodin}. A
single molecule is described at a coarse grained level with explicit account
of intramolecular interactions while the interactions between different
molecules are described through mean molecular fields which are found
self-consistently.

\begin{figure}[tbp]
\centerline{\includegraphics[width=.7\textwidth]{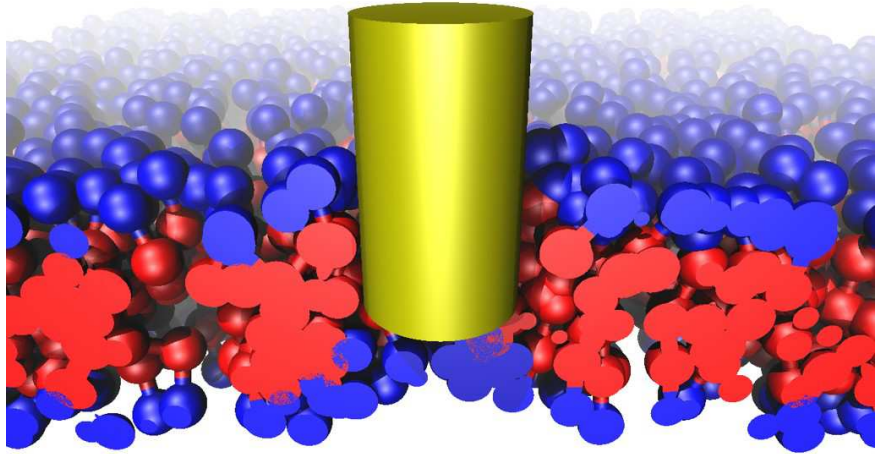}}
\caption{Mean field "snapshot" (set of most-probable conformations of the
lipids) of the perpendicular SWNT insertion produced by the SCMF theory.
Nanotube diameter 2.43 nm, nanotube position $-0.63$ nm, interaction
parameter between the nanotube and the hydrophobic core $\protect\varepsilon%
_T=-6.30$ kT.}
\label{Figure-01}
\end{figure}

In the present work the phospholipid molecule is modeled within the 3-beads
model with the interaction parameters obtained in Ref. \citenum{Pogodin}.
This model allows for spontaneous self-assembly of phospholipids in bilayers
with realistic bulk bilayer properties such as the average interfacial area
per lipid at equilibrium, the thickness of the bilayer and its hydrophobic
core, and the compressibility constant. These thermodynamic properties
rigidly fix the compression -- extension energy curve\cite{Pogodin}, the
output of the SCMF calculations: the minimum of this curve gives the
equilibrium area per lipid while its derivative is proportional to the
compressibility constant. Thus, this curve gives us confidence in calculated
values of the energy upon compression - extension of the bilayer.

The carbon nanotube is modeled as a rigid cylinder of a given diameter which
is oriented perpendicularly to the bilayer plane. More precisely, the carbon
nanotube represents a cylindrical region in the simulation box, which is not
accessible for phospholipids. We assume that the cylinder can interact with
hydrophobic tails of phospholipids, because the driving force for the
assembly of the phospholipid bilayer is the hydrophobic interactions and,
thus, the interactions between the hydrophobic tails lead to the main
contribution in the free energy.

The aim of this work is an accurate measurement of the energy cost of the
presence of the SWNT at a fixed position from the bilayer plane. Since the
SWNT can perturb the equilibrium structure of the bilayer, the free bilayer
may ripe around, bend or displace in order to minimize the perturbation. In
order to avoid such movements, the position of the bilayer in our model is
restricted by non-interacting walls at the top and the bottom of the
simulation box and the periodic boundary conditions are applied on the
sides. Since the reference state for the free energy calculation is the
energy of unperturbed phospholipid bilayer, we assume that the simulation
box represents a small part of a much larger system, where the part of the
bilayer outside the box is approximated by the unperturbed bilayer which
serves as a reference state. Thus, calculating the free energy of the small
box, containing a part of the lipid bilayer, we can construct the free
energy of the larger system. We write the total free energy of the system, $%
F $, as a sum of the free energy of the simulation box, $F_{box}$, and the
free energy of the system outside the box, $F_{out}$, which, in turn, is
expressed \textit{via} the free energy density of the unperturbed bilayer
given by the reference state, $f_{l}^{0}$, and the entropy of pure solvent, $%
f_{s}=\frac{\phi _{0}}{v_{s}}\ln \frac{\phi _{0}}{v_{s}}$, where $v_{s}$ is
the volume of the solvent molecule and $\phi _{0}$ is the bulk solvent
volume fraction (see Ref. \citenum{Pogodin}),

\begin{equation}
F=F_{box}+F_{out}=F_{box}+V_{l}^{out}f_{l}^{0}+V_{s}^{out}f_{s}
\end{equation}
The volume of the bilayer part lying outside of the simulation box, $%
V_{l}^{out}$, and the volume of pure solvent outside of the box, $%
V_{s}^{out} $, can be expressed \textit{via} the volume of the unperturbed
bilayer in the simulation box, $V_{l}^{0}$, and the equilibrium numbers of
lipids in the box with the cylinder, $N_{l}$, and without cylinder, $%
N_{l}^{0}$. In addition, we take into account the conservation of the total
volume of the system and the total number of lipid and solvent molecules.
With this, $f_{l}^{0}$ is related to the free energy $F_{box}^{0}$ of the
box containing the unperturbed bilayer as
\begin{equation}
f_{l}^{0}=\frac{1}{V_{l}^{0}}\left(
F_{box}^{0}-(V_{box}-V_{l}^{0})f_{s}\right)
\end{equation}
while the free energy difference with respect to the reference state of
unperturbed bilayer yields in the form
\begin{equation}
\Delta F=F_{box}-\frac{N_{l}}{N_{l}^{0}}\left(
F_{box}^{0}-V_{box}f_{s}\right) - \left(V_{box} -V_{cyl}^{in}\right)f_{s}
\end{equation}
where $V_{box}$ is the volume of the simulation box, and $V_{cyl}^{in}$ is
the part of the cylinder lying inside the box. We used 2D cylindrical
geometry in order to discretize the space such that the symmetry axis
coincide with the axis of the cylinder.

\section{Results and Discussion}

We have performed series of SCMF calculations for different positions of
perpendicularly oriented SWNT with respect to the DMPC phospholipid bilayer
plane for different diameters of the SWNT and interaction parameters between
the SWNT and the core of the bilayer. The output of the calculations is the
mean field concentration profiles of the heads and the tails in the bilayer,
demonstrating the structural rearrangements of phospholipids induced by the
SWNT at the molecular level and the precise measurements of the equilibrium
free energy change for each position of the SWNT. In addition, the SCMF
theory gives the probabilities of each phospholipid conformation in the
corresponding mean fields. This information allows for visualization of the
equilibrium molecular structure of the bilayer disturbed by the presence of
the SWNT in form of mean field "snapshots" representing the most probable
conformations of the molecules and their positions in the layer. One of such
representative "snapshots" is shown in \ref{Figure-01}.

\begin{figure}[tbp]
\begin{center}
\centerline{\includegraphics[width=.5\textwidth]{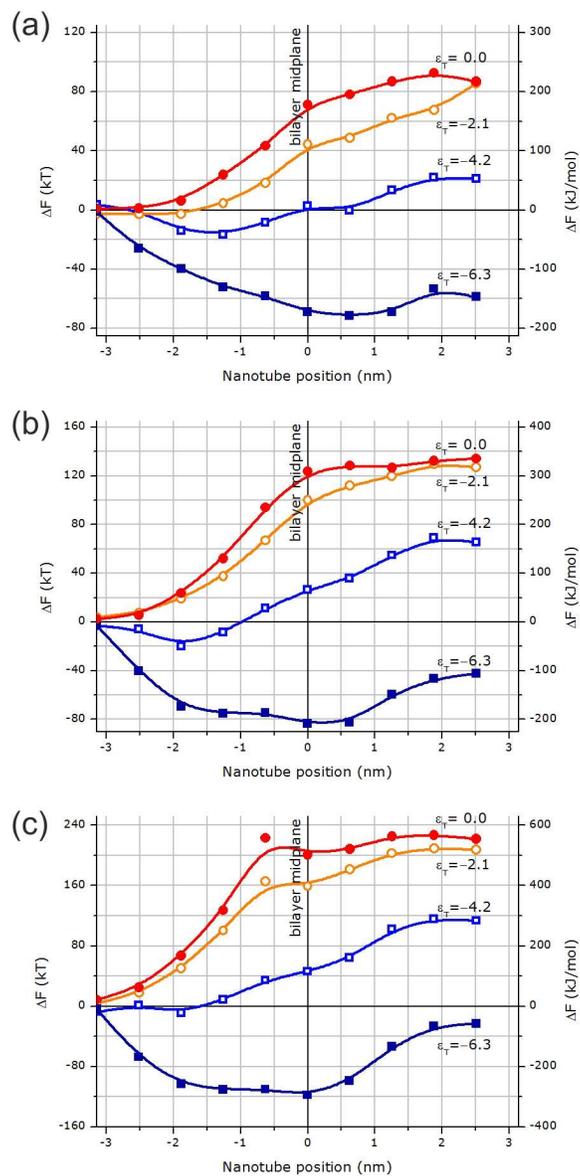}}
\end{center}
\caption{Free energy cost $\Delta F$ \textit{versus} nanotube position of
SWNTs with diameters 1 nm (a), 2.43 nm (b) and 4.86 nm (c) and different
interaction parameters with the hydrophobic core of the phospholipid
bilayer, $\protect\varepsilon_T$.}
\label{Figure-02}
\end{figure}

Carbon nanotubes, by their chemical structure, can be considered as
hydrophobic cylinders \cite{Bianco3}. Such objects in aqueous solutions tend
to aggregate in bundles, which make difficult to observe individual SWNT in
solution. In order to prevent such aggregation, the surface of the SWNT is
often treated with acids or functionalized \cite{Raffa,Bianco3} in such a
way that they become slightly hydrophilic. That is why, the energy per
contact of the hydrophobic bead with the nanotube, $\varepsilon_T$, in our
calculations changes from 0, representing steric repulsion, to $-6.3$ kT,
which corresponds to strong attraction (we assume the interaction distance
8.1 nm \cite{Pogodin}). Since SWNTs may have diameters ranging from 1 nm to
5 nm, we have chosen three diameters of the cylinder, 1.00 nm, 2.43 nm and
4.86 nm. The latter diameter is, probably, too big for SWNTs, but it may
correspond to multi-walled nanotubes and we include it for completeness.

The free energy cost of the equilibrium insertion of the SWNT as a function
of the distance between the SWNT tip and the middle plane of the unperturbed
layer for different interaction parameters is shown in \ref{Figure-02}. Here
the SWNT position $-3.14$ nm corresponds to the SWNT at the bilayer surface,
while the position $3.14$ nm corresponds to fully inserted SWNT. Insertion
of slightly hydrophilic SWNT is not favorable and the free energy cost
increases with the insertion distance until the SWNT pierce completely the
bilayer and the curve reaches the plateau, where the nanotube can slide
along its length through the bilayer with no energy cost. In turn, the
insertion of a hydrophobic SWNT is favorable and the corresponding insertion
curve has a pronounced minimum which corresponds to the partial insertion of
the nanotube into the bilayer when the hydrophobic bottom of the cylinder
has maximum contacts with the phospholipid tails. It is noteworthy, that
this minimum indicates that the perpendicular position of the hydrophobic
nanotube would be less favorable than the parallel insertion into the core
of the bilayer when the SWNT can have even larger number of contacts with
the phospholipid tails. The corresponding free energy would be even more
negative. The intermediate interaction parameters lead to mixed behavior,
the energy gain at a partial insertion and the energy lost at a full
insertion.

\begin{figure*}[tbp]
\begin{center}
\centerline{\includegraphics[width=1.0\textwidth]{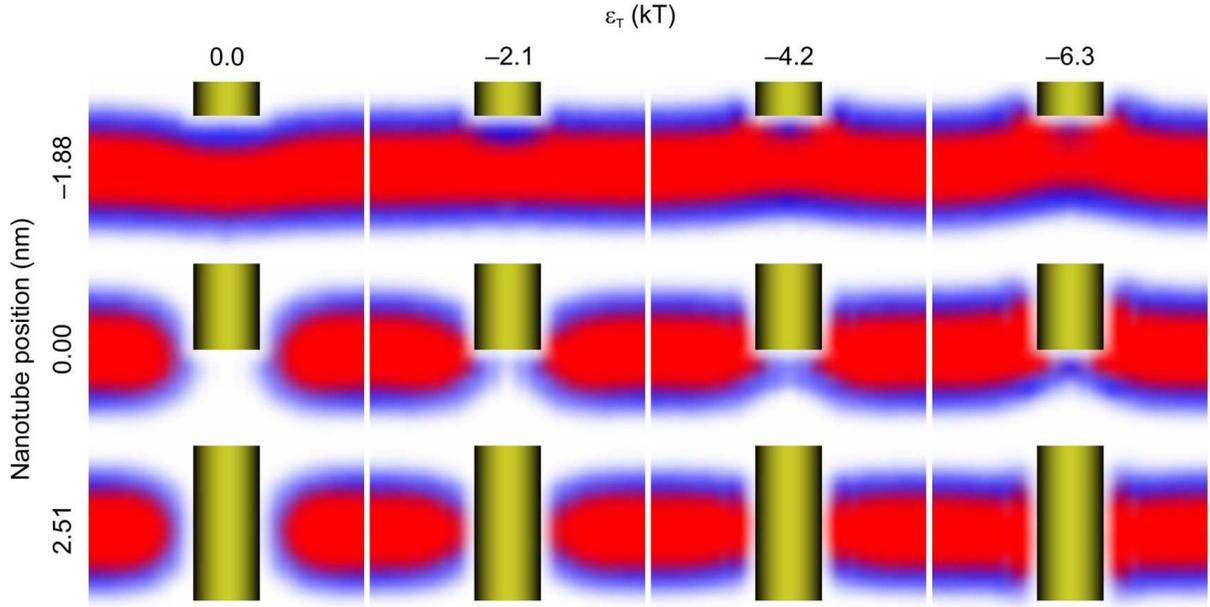}}
\end{center}
\caption{Morphology of the phospholipid bilayer structure induced by the
equilibrium insertion of 2.43 nm SWNT for different interaction parameters
between the nanotube and the hydrophobic core, $\protect\varepsilon_T$.}
\label{Figure-05}
\end{figure*}

Inserted nanotube induces structural changes in the phospholipid bilayer.
The SCMF theory allows to visualize the equilibrium structure of the bilayer
at each position of the nanotube \textit{via} equilibrium concentration
profiles (see \ref{Figure-05}). One can see how the increasing
hydrophobicity of the SWNT changes the character of the interaction with the
bilayer and the structural changes around the SWNT. In case of steric
repulsion ($\varepsilon_T = 0.0$ kT) the cylinder compresses the bilayer and
the tails tend to hide from the nanotube and the solvent, however
hydrophobic SWNTs ($\varepsilon_T = -2.1, -4.2, -6.3$ kT) attract
hydrophobic core of the layer which tend to stick to the edges of the
nanotube. The surface of the bilayer even lifts up in order to stick to the
nanotube and increase the area of contact between the SWNT and the
hydrophobic tails of the lipids. This is consistent with the corresponding
decrease of the free energy upon insertion of the SWNT observed in \ref%
{Figure-02}. Full insertion of the nanotube leads to the pore formation. The
energy of the pore formation as well as the structure of the pore strongly
depends on the hydrophobicity of the nanotube $\varepsilon_T$. In the most
hydrophilic case, the bilayer core is separated from the nanotube by the
layer of heads, while in the most hydrophobic case, $\varepsilon_T = -6.3$
kT, the tails touch the surface of the nanotube and one can observe the
wetting of the nanotube with a pronounced rim around the nanotube. \ref%
{Figure-05} provides also information about the breakthrough distances. The
hydrophilic nanotube creates a pore in the bilayer when it is inserted in
the middle of the bilayer, in turn, hydrophobic nanotubes have to cross
almost the full thickness of the bilayer before piercing it. This behavior
is consistent with a similar observation in AFM experiments of AFM tip
insertion in SOPC/cholesterol bilayers \cite{Melosh}.

\begin{table}[tbp]
\caption{Free energy change due to full insertion of the SWNT, $\Delta
F_{full}$, the minimum free energy, $\Delta F_{min}$, the insertion distance
corresponding to the minimum of the free energy, $d_{min}$ and the maximal
force to be applied to pierce the bilayer.}
\label{Table}{\scriptsize
\begin{tabular*}{\hsize}{@{\extracolsep{\fill}}r|rrrr|rrrr|rrrr}
\hline
Diameter & \multicolumn{4}{c|}{1.00 nm} & \multicolumn{4}{c|}{2.43 nm} &
\multicolumn{4}{c}{4.46 nm} \\ \hline
$\varepsilon_{T}$ (kT) & $0.0$ & $-2.1$ & $-4.2$ & $-6.3$ & $0.0$ & $-2.1$ &
$-4.2$ & $-6.3$ & $0.0$ & $-2.1$ & $-4.2$ & $-6.3$ \\ \hline\hline
$\Delta F_{full}$ (kT) & 87.3 & 85.9 & 21.2 & $-59.1$ & 134.5 & 127.4 & 65.6
& $-42.7$ & 222.8 & 208.1 & 113.7 & $-23.3$ \\ \hline
$\Delta F_{min}$ (kT) & 0.8 & $-3.0$ & $-16.7$ & $-72.1$ & 2.3 & 3.8 & $%
-20.3 $ & $-83.8$ & 8.6 & 3.5 & $-9.0$ & $-118.6$ \\ \hline
$d_{min}$ (nm) & $-3.14$ & $-2.51$ & $-1.25$ & 0.63 & $-3.14$ & $-3.14$ & $%
-1.88$ & 0.00 & $-3.14$ & $-3.14$ & $-1.88$ & 0.00 \\ \hline
Max. force (pN) & 157 & 132 & 74 & 61 & 237 & 206 & 116 & 120 & 517 & 381 &
184 & 241%
\end{tabular*}
}
\end{table}

The numerical values of the free energy cost of translocation through the
bilayer are summarized in \ref{Table}. The energy cost to cross the bilayer
is quite high for all diameters and interaction parameters (hundreds kT).
Hydrophilic nanotubes have a positive energy barrier, which may be difficult
to overcome by thermal motion, while the attraction between hydrophobic
nanotubes and phospholipid tails is high enough to entrap the nanotubes in
the core of the bilayer. Nanotubes with intermediate hydrophobicity exhibit
both effects, steric repulsion due to the pore formation and the enthalpic
attraction to the core of the bilayer. Thus, the nanotubes with intermediate
parameters will have to cross both barriers and the resulting energy cost is
the sum of the two barriers. Furthermore, the perpendicular orientation of
the nanotube has the lowest energy cost for penetration through the layer.
This orientation is not necessary the equilibrium one. For example, the
hydrophobic nanotube would preferentially orient itself parallel to the
layer in the hydrophobic core, where the nanotube would have more contacts
with the core. This orientation will have much lower energies, which are
proportional to its length. Thus, we would expect that hydrophobic nanotubes
with $\varepsilon_{T}=-4.2$ and $-6.3$ kT should stick in the hydrophobic
core in a parallel orientation. In turn, hydrophilic nanotubes in a parallel
orientation will have more steric contacts with both types of phospholipid
beads. Thus, the positive barrier would be higher. The translocation through
the bilayer also implies the diffusion of the nanotube in a correct
orientation close to the bilayer. This would require additional time, thus
effectively increasing the energy barrier.

High values of the energy barrier can be understood with the help of a
simple estimation. If we consider that the cylinder with the radius $%
R_{cyl}=1.2$ nm simply moves apart the lipids and creates a pore with the
radius $R_{cyl}+R_{bead}$, where $R_{bead}=0.4$ nm is the radius of the bead
of in the phospholipid molecule, the pore formation energy is due to
breaking contacts between phospholipids lying close to the edge of the pore.
Taking into account that the perimeter of the pore is $%
2\pi(R_{cyl}+R_{bead})\sim 10$ nm, the area per lipid is 60 \AA $^2$, the
energy per lipid due to tail-tail contacts in the 3-beads model is $\sim -20$
kT and assuming that lipids loose half of the contacts and the membrane is a
double layer, we find that the number of lipids at the edge is $\sim 26$,
\textit{i.e.} the energy cost for the formation of the pore is $\sim 260$
kT, which is comparable by the order of magnitude but higher than the
calculated value of 192 kT. The discrepancy is due to contributions of the
heads and the rearrangements of the lipids around the pore in order to
minimize the energy cost, which does not have a sharp edge. Note, that this
energy is comparable with the energy of breaking a chemical bond since
typical values of bond energies are also hundreds kT \cite{handbook}.

The calculated equilibrium force of hundreds pN corresponds to the energy of
hundreds kT, which is extremely high energy at the molecular level. However,
the nanotube can pierce the phospholipid layer if the external force is
applied. Carbon nanotubes can be fixed on AFM tip and they can be used as
"nano-injectors" or nano-probes \cite{Macpherson}. Such combination of AFM
and carbon nanotube can be used for force measurements. The measurements on
living cells \cite{Moudgil} have shown that the force of penetration depends
on the location over the cell membrane and, possibly, many other factors,
especially if the cytoskeleton is involved in the response to the stimulus.
The minimal force for piercing of cell membrane measured in these
experiments is of order 100 pN, \textit{i.e.} even higher than suggested by
our calculations.

The strength of our method is the direct and accurate calculation of free
energies at equilibrium, that allow for judgement about the possibility of
spontaneous translocation through the bilayer. The forces obtained in
non-equilibrium MD simulations of nanotube penetration with a constant speed
\cite{Sansom1} are several times higher than in our calculations and they
depend on the speed of pulling, since the layer needs some time to
accommodate its structure to the external perturbation. Hence, these
calculations at low speed can represent an upper limit for the equilibrium
free energy. In contrast, the SCMF method gives directly the equilibrium and
average picture which does not show the positions of individual
phospholipids but the average concentration profiles. That is why our
equilibrium mean field "snapshots" and average concentration profiles look
different from the simulation snapshots of non-equilibrium piercing \cite%
{Sansom1}.

The results with different diameters of nanotubes suggest that the thinner
is a nanotube, the less is the energy barrier. Thus, we may conclude that
larger objects such as multi-walled carbon nanotubes composed of several
graphitic concentric layers \cite{Bianco3} and having diameters more than 4
nm will have even larger energy barrier, which require the application of an
external force to pierce through the bilayer. Finally, an illustration of
strong resistance of phospholipid bilayers against piercing is provided in
the experiments with microtubules (rigid cylinders of 30 nm in diameter)
growing inside the phospholipid vesicles \cite{Elbaum}. Growing microtubule
exert force on the inner wall of the vesicle at both ends, which leads to
large deformation of the surface of the vesicle and the buckling of the
microtubule.

\section{Conclusions}

The SCMF theory is used for accurate measurements of the energy cost
associated with the perpendicular insertion of carbon nanotubes of different
diameters in the phospholipid bilayer. This method gives direct access to
the equilibrium free energy and provides microscopic information about the
structural rearrangements of the phospholipid bilayer around the inserted
nanotubes. This information, though obtained for model DMPC phospholipid
bilayers, may contribute to the discussion of the possible mechanisms of
translocation of SWNTs into cells. In particular, it reflects on the
inferred mechanism of spontaneous translocation of SWNTs through
phospholipid bilayers by thermal motion.

Our results suggest that hydrophilic or non-interacting SWNTs would require
the energy of order hundreds kT to cross the phospholipid bilayer in
perpendicular orientation. Furthermore, perpendicular orientation is less
disruptive for the bilayer and thus, other orientations would require even
higher energies. In turn, more hydrophobic, interacting SWNTs are attracted
by the hydrophobic cores, what hinders their translocation through the
bilayers and renders difficult the separation of the nanotube from the
bilayer simply by thermal motion. Thus, our results for model phospholipid
bilayers may suggest that experimentally observed translocation of SWNTs
into cells is probably due to other \textit{energy dependent} translocation
mechanisms such as, for example, endocytosis.

These results for homogeneous nanotubes may also indicate the ways to reduce
the energy cost of translocation through the phospholipid bilayers. For
example, inhomogeneous patterning of the nanotube surface can help in design
of nano-objects which can freely pass through or preferentially associate
with the phospholipid bilayers.

The predicted energy of the bilayer rupture due to the perpendicular
insertion of a carbon nanotube can be directly verified by the force
measurements of the insertion of the AFM tip functionalized with a carbon
nanotube.

\section{Method: Single Chain Mean Field theory}

The SCMF theory describes a single molecule surrounded by the mean fields
\cite{Ben-Shaul1,Ben-Shaul2,Bonet1,Pogodin}. It takes explicitly into
account the structure of an individual molecule at a coarse grained level
similar to coarse grained MC or MD simulations. However, as distinct from
simulations, the interactions of the molecule with the environment are
described through the mean molecular fields. The mean fields determine the
most probable conformations of the molecule through the probabilities of
individual molecule. In turn, the mean fields are calculated as the average
properties of individual conformations. This self-consistence closure
defines the set of non-linear equations that can be solved numerically. The
solution of such equations gives the equilibrium structures and the
concentration profiles of all components in the system as well as the most
probable conformations of individual molecules.

To model the phospholipid bilayer we use the 3-beads model of the
DMPC phospholipid molecule described in Ref. \citenum{Pogodin}.
The phospholipid molecule is modeled as three spherical beads of
diameter $d= 0.81$ nm, joined consequently by stiff bonds of 1.0
nm in length: two hydrophobic, T-beads, representing the tails,
and one hydrophilic, H-bead, representing the phospholipid head.
The angle between the bonds is free to bend as long as the
terminal beads do not intersect each other. The solvent molecule
is modeled as a S-bead of the same diameter as the phospholipid
beads. The interactions between the beads in this model are given
only by two square well potentials with interaction range
$r_{int}=1.215$ nm and depths $\varepsilon _{TT}=-2.10$ kT for T-T
contacts and $\varepsilon _{HS}=-0.15$ kT for H-S contacts.
Additional energy is assigned to the T-beads residing at a
distance shorter than 8.10 nm from the surface of the cylinder,
$\varepsilon _{T}$ (see \ref{Table}). The simulation box
$10.00\times 10.00\times 6.27$ nm ($12.00\times 12.00\times 6.27$
nm for the case of the largest nanotube's diameter) is discretized
into a grid according to 2D cylindrical geometry with the symmetry
axis placed in the center of the box and oriented along the $z$
-axis. The grid cells are about 0.5 nm in the radial and 0.3 nm in
the vertical directions (comparable with the diameters of the
beads in the 3-beads model, 0.81 nm) and the sampling of the
molecule conformational space is 400 000 conformations. A series
of test calculations have shown that these parameters give enough
precision for the energy calculations.

We use the generalized equations of the SCMF method provided in Ref. %
\citenum{Pogodin} which describes the self-assembly of a mixture of an
arbitrary number of types of molecules of an arbitrary structure. First step
of the SCMF method is the generation of the representative sampling $%
\{\Gamma \}$ of conformations of a single molecule, where the conformations
differ in position and orientation of the molecule in the space and in the
angle between the bonds of the molecule. If the sampling was generated with
a bias, it is corrected with a known weight of each conformation $w(\Gamma )$%
. The probability of a given conformation $\Gamma $ of the molecule, placed
in the simulation box, containing $N$ lipids is given by\cite{Pogodin}

\begin{equation}
\rho (\Gamma )=\frac{1}{Zw(\Gamma )}\exp \left( -H^{intra}(\Gamma
)-(N-1)\sum_{i}\varepsilon _{i}^{T}(\Gamma )c_{i}^{T}-\sum_{i}\varepsilon
_{i}^{S}(\Gamma )\frac{\phi _{i}^{S}}{v_{s}}+\sum_{i}V_{i}\lambda _{i}\phi
_{i}(\Gamma )\right)   \label{eqforrho}
\end{equation}%
where $Z$ is the normalization constant, $H^{intra}(\Gamma )$ is
the internal energy of the conformation $\Gamma $, where we
include, as well, its interaction with the nanotube. The
simulation box is discretized into a grid cells $i$ of the volume
$V_{i}$. The interactions of the beads with the fields
($c_{i}^{T}$ and $c_{i}^{H}$ are the equilibrium concentrations of
the beads of each type,) are described through interaction fields
of a given conformation, $\varepsilon _{i}^{T}(\Gamma)$ and
$\varepsilon _{i}^{S}(\Gamma)$. The Lagrange multiplier $\lambda
_{i}$ is related to the total volume fraction
of the cell $i$ occupied by the solvent, $\phi _{i}^{S}$ \textit{via} $%
v_{s}\lambda _{i}=\ln \phi _{i}^{S}+N\varepsilon _{i}^{H}c_{i}^{H}$, where $%
v_{s}$ is the volume of the solvent molecule and $\varepsilon
_{i}^{H}=\frac{4}{3}\pi \left( r_{int}^{3}-d^{3}\right)
\varepsilon_{HS}$. The condition of local incompressibility
relate $\phi _{i}^{S}$ with the volume fraction of the lipids
$\phi _{i}$ \textit{via} $\phi _{i}^{S}+N\phi _{i}=\phi _{0}$,
where $\phi _{0}$ is the maximal volume fraction occupied by the
solvent and all the molecules (equal to 0.675 in this model) \cite{Pogodin}.

Once the probabilities of the conformations $\rho (\Gamma )$ are known, the
molecular mean fields, namely the volume fractions of the molecules and
concentrations of the beads of all types, are calculated as averages over
all conformations,
\begin{equation}
\phi _{i} =\sum_{\{\Gamma \}}\phi _{i}(\Gamma )\rho (\Gamma )  \label{eqn}
\end{equation}
\begin{equation}
c_{i}^{T(H)} =\sum_{\{\Gamma \}}c_{i}^{T(H)}(\Gamma )\rho (\Gamma )
\label{eqn2}
\end{equation}

These equations, \ref{eqforrho}, \ref{eqn}, \ref{eqn2}, define the closed
set of nonlinear equations for the probabilities of conformations and the
mean fields. Solution of these equations gives the equilibrium properties of
the system and the equilibrium free energy of the simulation box, which is
written \cite{Pogodin} in units kT as
\begin{eqnarray}
F_{box} &=&N\left\langle \ln N\rho (\Gamma )w(\Gamma )\kappa \right\rangle
+\sum_{i}V_{i}\frac{\phi _{i}^{S}}{v_{s}}\ln \frac{\phi _{i}^{S}}{v_{s}}%
+N\left\langle H^{intra}(\Gamma )\right\rangle   \notag \\
&&+\frac{1}{2}N(N-1)\sum_{i}\left\langle \varepsilon _{i}^{T}(\Gamma
)\right\rangle c_{i}^{T}+N\sum_{i}\left\langle \varepsilon _{i}^{S}(\Gamma
)\right\rangle \frac{\phi _{i}^{S}}{v_{s}}
\end{eqnarray}%
where $\kappa $ is the number of conformations in the sampling $\{\Gamma \}$
and all the averages, denoted by angular brackets are taken as averages over
the sampling with probabilities $\rho (\Gamma )$ in analogy to \ref{eqn} and %
\ref{eqn2}.

The calculation time of one point in the energy curve and full set of
microscopic data of the bilayer configuration takes less than 1 day on a
single 32-core machine. This time can be drastically decreased by taking
advantage of high capability of parallelization of the SCMF method, which
allows to use more computers and processors.

\begin{acknowledgement}

The authors wish to thank Prof. Nigel Slater from Cambridge
University for useful discussions, suggestions, comments and
collaboration within UK Royal Society International Joint Project
with Cambridge University. We acknowledge the financial help from
Spanish Ministry of education MICINN \textit{via} the project
CTQ2008-06469/PPQ.

\end{acknowledgement}

\begin{suppinfo}

Structural changes in the phospholipid bilayer induced by the
equilibrium insertion of a carbon nanotube for different distances
of a nanotube from the bilayer midplane.

\end{suppinfo}


\providecommand*\mcitethebibliography{\thebibliography} \csname
@ifundefined\endcsname{endmcitethebibliography}  {\let%
\endmcitethebibliography\endthebibliography}{}

\begin{tocentry}

\begin{center}
\includegraphics[width=8.5cm]{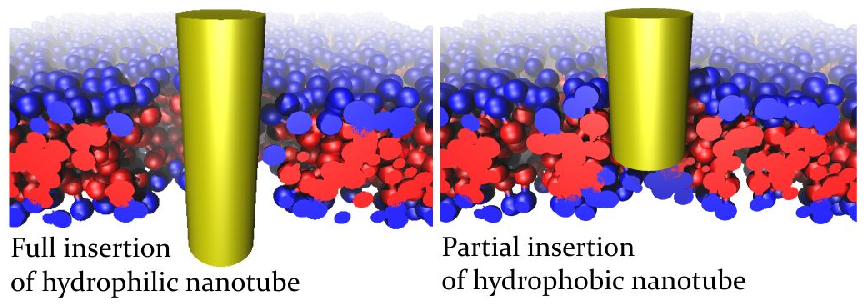}

\end{center}
\end{tocentry}

\end{document}